\documentclass[a4paper,11pt]{article}
\pdfoutput=1 % if your are submitting a pdflatex (i.e. if you have
\usepackage{jinstpub} % for details on the use of the package, please
\usepackage{enumitem}
\usepackage{url}
\usepackage{rotating}
\usepackage{hyperref}

\title{Design of a Helium Passivation System for the Target Vessel of the Beam Dump Facility at CERN}

\author[a,1]{P.~Avigni,\note{Corresponding author.}}
\author[a]{M.~Battistin,}
\author[a]{M.~Calviani,}
\author[b]{P.~Dalakov,}
\author[a]{K.~Kershaw,}
\author[b]{J.~Klier,}
\author[a]{M.~Lamont,}
\author[a]{E.~Lopez~Sola}
\author[a]{and~J.~M.~Martin~Ruiz,}

\affiliation[a]{CERN, 1211 Geneva 23, Switzerland}
\affiliation[b]{Institut f\"{u}r Luft- und K\"{a}ltetechnik gemeinn\"{u}tzige Gesellschaft mbH (ILK Dresden), Bertolt-Brecht-Allee 20, Dresden, Germany}

\emailAdd{pietro.avigni@cern.ch}

\abstract{
    The Beam Dump Facility (BDF) is a proposed general-purpose facility at CERN, dedicated to fixed target and beam dump experiments, currently being developed in the context of the Physics Beyond Colliders program. The design of the facility will allow to host different types of experiments, of which SHiP is planned to be the initial one. The core of the facility is a high-density target/dump absorbing the full intensity of the SPS beam and generating a cascade of particles that are detected downstream the target complex. The target and its shielding blocks are positioned inside a vessel, which is planned to be passivized with helium, in order to reduce the activation of the gas surrounding the target and to extend the operational life of materials and equipment. The passivation system that will be in charge of purifying and circulating the helium is a critical component for the operation of the facility. Fluid dynamics simulations have been performed to study the circulation of the helium through the vessel. A detailed design of the helium passivation system and its main components has been developed.
    }


\arxivnumber{1910.00333} % only if you have one

\begin{document}
\maketitle
\flushbottom

\section{Introduction}
\label{Sec:Intro}

The Beam Dump Facility (BDF) is a proposed general-purpose facility for high-intensity fixed target experiments that is currently being developed at European Laboratory for Particle Physics (CERN). The facility is planned to be built in the North Area in the Prevessin campus and it is designed with sufficient flexibility to be reused for different experimental setups. The initial experiment that is planned to be installed in the area is the SHiP (Search for Hidden Particles) experiment~\cite{Anelli_SHiP,Alekhin_SHiP}; this experiment is based on a dense target~\cite{PhysRevAccelBeams.22.113001}, absorbing the full intensity of the SPS beam and maximizing the production of feebly interacting particles.

\begin{figure}
\centering
\resizebox{0.85\textwidth}{!}{
\includegraphics{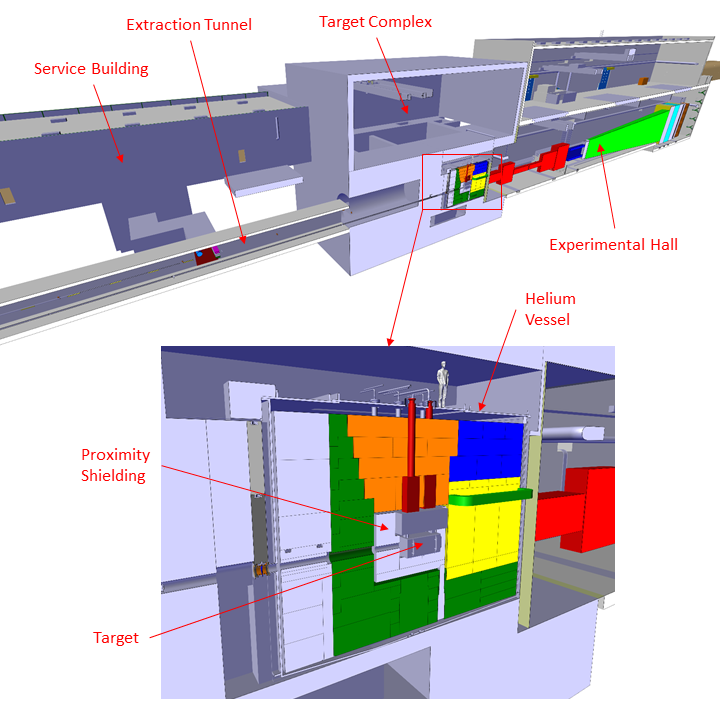}}
\caption{\label{Fig:Intro:HeVeLayout} Cross-section view of the target complex, the helium vessel and its internal components on a vertical plane containing the beam axis.}
\end{figure}

The target is located in a dedicated building, the target complex. The target is a critical component of the facility, due to the high power deposition level (roughly 300 kW on target, plus 50 kW on surrounding shielding~\cite{PhysRevAccelBeams.22.113001}) and an expected operational life of five years. The target complex is followed downstream by an experimental hall hosting a series of magnets and detectors specific to the installed experimental setup. 
The target is surrounded by thick shielding blocks to minimize radiation dose to the surrounding environment. The shielding blocks with the target are located inside a vessel, which is planned to be filled with helium gas. Baseline parameters of the helium vessel are provided in Table~\ref{Tab:Intro:HeVeParm} and a vertical cross section is illustrated in Figure~\ref{Fig:Intro:HeVeLayout}.

Before starting the operation of the target, air inside the vessel is flushed and replaced by helium; during the operation, the helium will be actively circulated and purified. Purified helium inside the vessel is needed for two main purposes:
\begin{itemize}
    \item Reduced activation: helium has a lower activation cross-section compared to air. In case of release of gas contained inside the helium vessel, the potential released contamination is substantially lower than air;
    \item Increased life of materials: since helium is an inert gas, specific reactions (such as oxidation) are prevented by design, thus increasing the lifetime of materials and equipment.
\end{itemize}

The helium is maintained in overpressure (+50 Pa) with respect to the surrounding environment, in order to prevent air from leaking into the vessel during operation. The surrounding environment is maintained in underpressure (-100 Pa) with respect to atmospheric pressure, in order to prevent activated air from being released to the environment without proper filtration by the target complex ventilation system. As a consequence, the pressure inside the vessel is normally -50 Pa with respect to atmospheric pressure.

One way of obtaining pure helium atmosphere inside the vessel is to pump air out of the vessel and then fill it with pure helium gas. However, this approach requires the vessel walls to be able to withstand the negative pressure differential generated by the vacuum. This aspect is particularly challenging for the BDF vessel, due to its large size. This technology is the one used in the T2K neutrino beam line at the J-PARC research center in Japan~\cite{ABE2011106}. A second drawback of the mentioned approach is the lack of purification capability: whenever a leak is detected and the helium is contaminated, the entire vessel needs to be purged and refilled with pure helium.

An alternative approach consists in flushing air out by supplying pure helium, lighter than air, from the top of the vessel and subsequently purifying the gas mixture up to the desired purity level. In order to deploy this approach, a helium passivation system needs to be included as part of the target complex design. 
This approach is superior in terms of structural requirements for the vessel and management of impurities inside the helium stream; however, it requires a dedicated helium passivation system integrated in the target complex to actively circulate, purify and cool the helium. The latter solution is selected for the preliminary design of the CERN's Beam Dump Facility because of the flexibility that it provides. 

The viability of the solution is dependent on the capability of properly circulating air and helium through the gaps between the shielding blocks and equipment inside the helium vessel.
A preliminary characterization of the helium flushing and circulation through the vessel has been performed via Computational Fluid Dynamics (CFD) simulation and it is presented in section~\ref{Sec:CFD}. Section~\ref{Sec:PSDes} presents the design of the helium passivation system that is responsible of circulating, cooling and purifying the helium in the vessel.

\begin{table}[hbtp]
\centering
\caption{\label{Tab:Intro:HeVeParm}
Baseline parameters of the BDF helium vessel}
\begin{tabular}{|l|c|}
\hline
Length {[}m{]}              	&    11.4                        \\
Width {[}m{]}            		&     8.5                 \\
Height {[}m{]}               	&    7.8                        \\
Free volume {[}m$^3${]}                  & 75                   \\
Pressure {[}Pa{]} relative to external volume            &   +50                  \\
Pressure {[}Pa{]} of external volume relative to atmosphere             &   -100                  \\
Average beam power  {[}kW{]}    & 356                       \\
Average beam power deposited on target {[}kW{]}    & 305                        \\
\hline
\end{tabular}
\end{table}

\section{Computational Fluid Dynamics of Helium Circulation and Flushing}
\label{Sec:CFD}

A CFD analysis of the flow characteristics inside the helium vessel has been performed in support of the BDF target complex design effort. The helium vessel contains a large number of components and equipment that can potentially constitute an obstacle against proper circulation of helium. The purpose of the analysis is to demonstrate that helium and air circulate properly through the available free volume and impurities can be efficiently extracted. In particular, the analysis has been conducted with the aim of demonstrating that the following two conditions are achieved:
\renewcommand{\labelenumi}{\Alph{enumi}.}
\begin{enumerate}
\item The formation of localized air pockets during initial flushing phase is minimal. This condition is fundamental in order to achieve a shorter and more efficient startup purification phase. This condition is favoured by the absence of obstructions in the downward flow path of air. A better extraction of air from the vessel also helps to reduce the amount of pure helium needed for the flushing phase;
\item The formation of stagnant gas areas during circulation phase is minimal. This condition helps to minimize the amount of impurities trapped inside the helium vessel that otherwise could not be adsorbed by the purification unit. As a consequence, the average activation of the gas mixture inside the helium vessel is reduced.
\end{enumerate}

In order to study the mentioned conditions, two sets of simulations have been run: 
\begin{itemize}
    \item Transient case simulating the start-up of the system via flushing. The simulation aims at identifying areas where purity of helium cannot be reduced below a specific threshold by limited-time flushing;
    \item Steady-state case simulating the nominal helium circulation through the vessel. The simulation aims at identifying areas where pure helium is stagnating (i.e. flow velocity is small). A limitation of the steady simulation approach is its inability for capturing impurity stagnation in recirculation areas, for which stagnation is not necessarily caused by low flow velocity; however, recirculation is captured in the transient case simulation, thus compensating for the limitations of the steady simulation. 
\end{itemize}
The steady-state single phase simulation for circulation is substantially less demanding in terms of computational resources, and it is presented in section~\ref{Sec:CFD:Circulation}, whereas the transient simulation is presented in section~\ref{Sec:CFD:Flushing}.
The solution of the CFD model includes the energy equation, beyond continuity and momentum equations. Due to the preliminary advancement level of the design of the helium vessel and its internal components, for which dedicated cooling systems are being designed, the CFD simulations presented in section~\ref{Sec:CFD:Circulation} and \ref{Sec:CFD:Flushing} have been run without heat loads. On the contrary, the design of the helium passivation system, presented in section~\ref{Sec:PSDes}, takes into account, for sizing purposes, the potential maximum heat load that might be present in the final installation.

\subsection{Model Preparation and Meshing}
\label{Sec:CFD:ModMesh}

A 3D model of the helium vessel and its internal equipment has been prepared in SpaceClaim~\cite{ANSYS_SpaceClaim}, based on the integration model for the BDF target complex~\cite{BDFcomplex}. Figure~\ref{Fig:Intro:HeVeLayout} shows a slice of the 3D model for the helium vessel, on the plane containing the beam axis and the vertical direction. The figure shows the target and the shielding blocks surrounding it, contained in the vessel walls, which are reinforced with a series of ribs in order to withstand the pressure differential. The packing of equipment inside the helium vessel is dense (shielding and target assemblies) and the free volume inside it is roughly 75 m$^3$ (substantially smaller than the overall vessel volume, which is on the order of 10$^3$ m$^3$).

Purified gas is supplied at the top of the helium vessel and exhaust gas is extracted at the bottom. In order to simplify the design of the vessel and the operations that need to be performed during maintenance, a single inlet penetration is foreseen for the supply as well as for the extraction; the upper and lower plenums distribute the flow and provide mixing.

\begin{figure}
\centering
\resizebox{0.85\textwidth}{!}{
\includegraphics{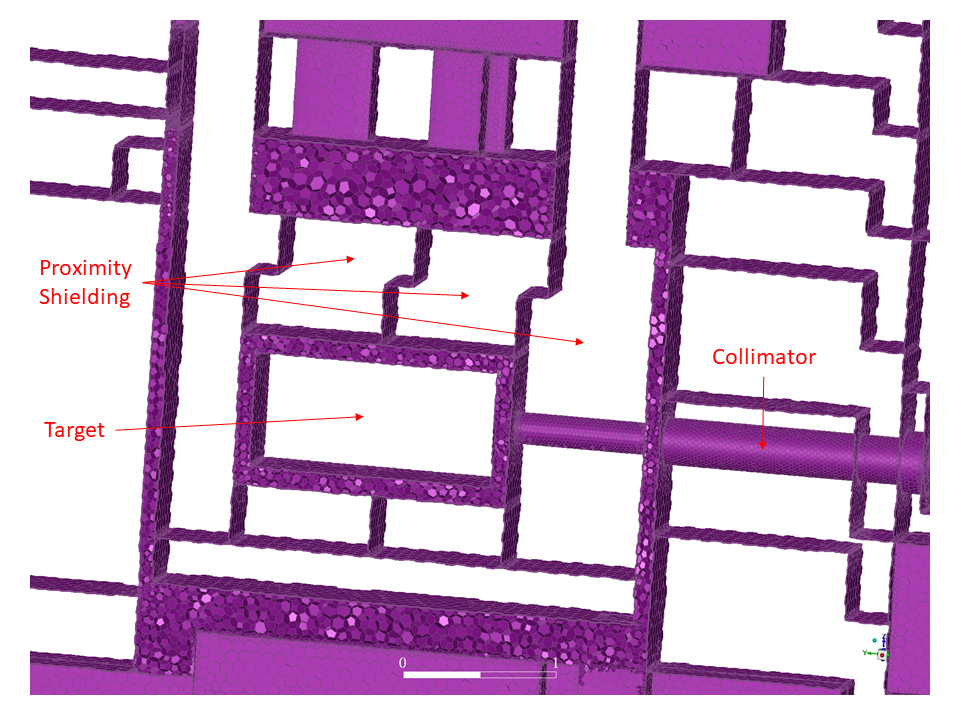}}
\caption{\label{Fig:CFD:Mesh} Mesh slice in the collimator-target region, showing the gaps between the target and the shielding blocks.}
\end{figure}

In order to facilitate the circulation of gas through the vessel, all shielding blocks are spaced by 1-cm-wide gaps. In terms of meshing, these gaps require high aspect-ratio cells; the presence of a large number of gaps and the overall size of the problem (7 million cells) make the meshing process quite challenging. The mesh has been prepared in ANSYS Fluent Meshing~\cite{ANSYS_FluentMeshing}, via the surface mesh approach and the prisms creation tools. The gaps have not been explicitly modeled in the 3D model; instead, they have been created using prism growth meshing, which allows to obtain high aspect-ratio cells with specific number of layers and thickness starting from a surface mesh. 
The mixed-type mesh has been converted to polyhedral mesh before simulation. 
Figure~\ref{Fig:CFD:Mesh} shows a mesh slice in the collimator-target region. The volumetric regions around the target are visible, as well as several thin gaps between blocks. Table~\ref{Tab:CFD:MeshStat} shows the mesh statistics; the mesh is around 7 million cells in size, the minimum and average orthogonal quality are 3.3\% and 29.4\%, respectively. The minimum quality required by Fluent for running the simulation is above 1\%, so the values used for the work are acceptable and the mesh check is passed according to Fluent standard rules. However the overall and minimum quality used for this simulation could be improved; this would allow to achieve a more accurate modeling of the flow through the gaps, reduced numerical diffusion (due to refined mesh) and thus a more precise evaluation of the time required to complete the flushing of the helium vessel. However, due to the limited computational capabilities available, an improvement of the orthogonal quality achieved via mesh refinement was deemed not feasible. Future more accurate analyses will be performed in order to improve the results in this direction.

\begin{table}[htbp]
\centering
\caption{\label{Tab:CFD:MeshStat} Statistics for the helium vessel mesh. A compromise between mesh quality and size has been achieved, in order to reduce the computational time of the simulations}
\smallskip
\begin{tabular}{|l|r|l|}
\hline
\textbf{Parameter} & \textbf{Value} & \textbf{Unit}\\
\hline
Size & 6782628 & cells \\
Allocated memory & 10982 & Mbytes \\
Total volume & 74.5 & m$^3$ \\
Min. orthogonal quality & 3.34 & \%\\
Max. aspect-ratio & 328 & -\\
Mesh check & Passed & -\\
\hline
\end{tabular}
\end{table}

\subsection{Helium Circulation}
\label{Sec:CFD:Circulation}

During nominal operation, the helium is circulated through the helium vessel via the blower that is included as part of the helium passivation system.

A steady-state simulation of the circulation of helium through the vessel has been run to characterize the nominal pressure and velocity distributions and to identify locations where the helium velocity is particularly small, leading to the formation of air pockets.

During cooling mode, a flow-rate of roughly 770 m$^3$/h is supplied by the blower; this flow-rate value is ten times larger than the purification flow-rate. The primary intent of the steady simulation is the identification of areas where the flow velocity is close to 0 m/s (stagnant flow). In order to identify stagnant flow areas, a threshold at 0.1 mm/s has been selected. The reason for the selection of this value as the threshold for identifying low-speed areas is the fact that higher values do not allow a clear identification of low-speed areas; in other words, at higher velocities the flow is more uniformly distributed. This empirical criterion is not necessarily ideal, but it allows a simple identification of critical areas; moreover, the areas identified using this criterion are consistent with the areas identified in the helium flushing section (section \ref{Sec:CFD:Flushing}). Future work will address more in detail the steady conditions, particularly the analysis of cases with steady impurity sources.

\begin{figure}
\centering
\resizebox{0.9\textwidth}{!}{
\includegraphics{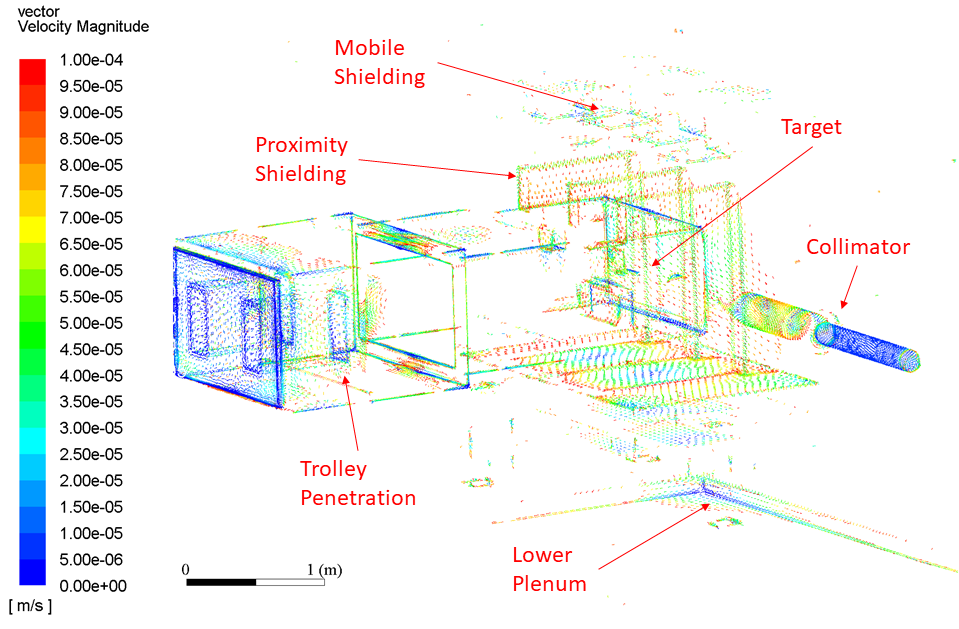}}
\caption{\label{Fig:CFD:AirPockets} Vector plot colored by velocity magnitude and clipped to 0.1 mm/s, showing regions of low flow velocity}
\end{figure}

In steady conditions, the velocity is generally not higher than 10 cm/s, except for the inlet and outlet plenums; the flow is distributed uniformly on the horizontal plane even if the pathlines between blocks are naturally quite irregular (due to changes of direction, corners, etc.). Figure~\ref{Fig:CFD:AirPockets} shows a vector plot colored by velocity magnitude and clipped to 0.1 mm/s, corresponding to 6 mm/min (areas of low velocity). The figure shows that the main critical areas for pocket formation are the trolley, the target, the proximity shielding and the collimator assembly. Collimator and trolley areas are less critical, due to the fact that the gaps are not excessively thin. The areas identified in the upper and lower plenum are not as critical because they are located in open volumes, for which recirculation and gravity provide a minimum level of mixing. With respect to the identified critical areas, the flow velocity is low, but flowpaths are provided in the design of the mentioned components. At the current design level, major changes to the design of the proximity shielding and target containment geometries are not planned. However, based on the analysis performed, it is recommended that the blocks and target are positioned precisely, in order to make sure that a 1 cm gap is present between them; dedicated spacers and grooves in the blocks are also foreseen in order to provide enhanced positioning precision and sufficient free volume for the helium flow.

\subsection{Helium Flushing}
\label{Sec:CFD:Flushing}

The circulation and purification of helium flowing through the vessel is achieved via the helium passivation system; this system is in charge of circulating and purifying helium during nominal operation. The purification function of the passivation system cannot be activated if the purity of the helium to be purified is lower than 85\%, in order to prevent condensation and freezing of air components at cryogenic temperatures. The mentioned constraint requires to flush the air in the vessel at every system startup, and replace it with helium up to a minimum purity of 85\%; the flushing is performed via the helium passivation system.

A transient simulation has been run to simulate the replacement of air with helium during the flushing process and identify air pockets remaining inside the helium vessel before purification is started. The setup of the simulation, in terms of 3D model and meshing, is analogous to the steady simulation. The fluid is a mixture of air and helium; the vessel is initially entirely filled with air, and at the inlet (top of helium vessel) pure helium is injected at a flow-rate of 770 m$^3$/h.
Since the free volume inside the helium vessel is roughly 75 m$^3$, at the nominal flow-rate, approximately 6 minutes are needed to perform 1 volume change; The simulation has been run for few volume changes.

\begin{figure}
\centering
\resizebox{0.95\textwidth}{!}{
\includegraphics{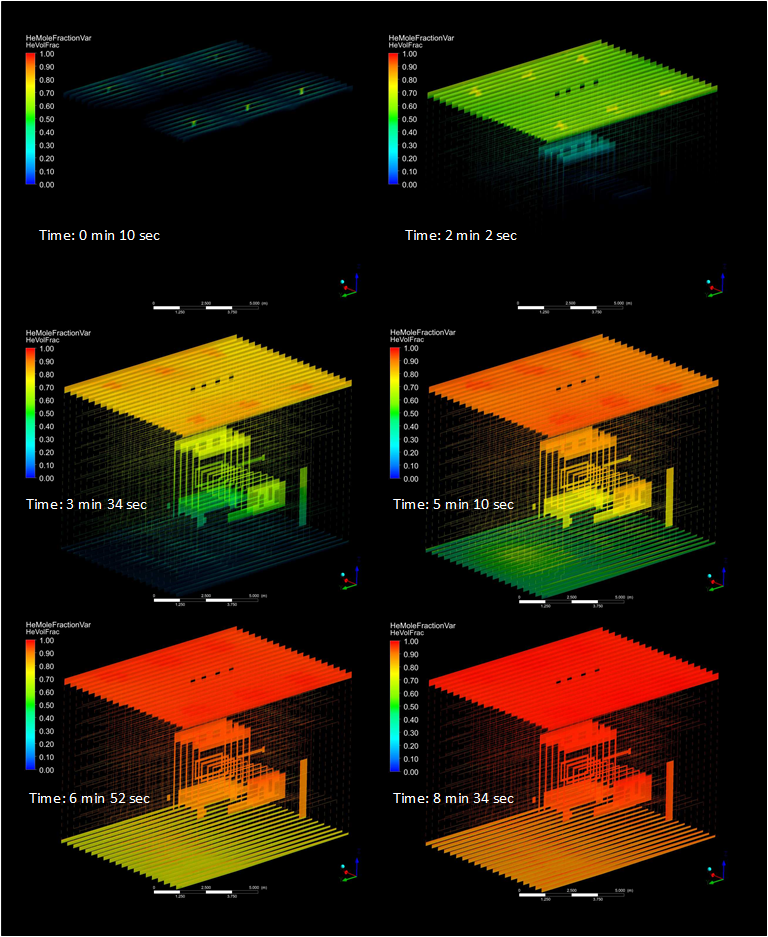}}
\caption{\label{Fig:CFD:FlushCont} Volume rendering of helium mole fraction during the flushing, showing the gradual filling of the vessel with helium pumped from the top inlet plenum}
\end{figure}

Figure~\ref{Fig:CFD:FlushCont} shows a volume rendering of helium purity (helium mole fraction) at six different instants of time during the simulation. As a reference, the last snapshot is taken at 8 min and 34 s after the beginning of the flushing, which is equivalent to approximately 1.4 volume changes. 

The series of snapshots shows that the helium gradually and uniformly fills the vessel, including its gaps, from top to bottom. After 1.4 volume changes, the average purity inside the vessel is consistently above 80\%.

\begin{figure}
\centering
\resizebox{0.8\textwidth}{!}{
\includegraphics{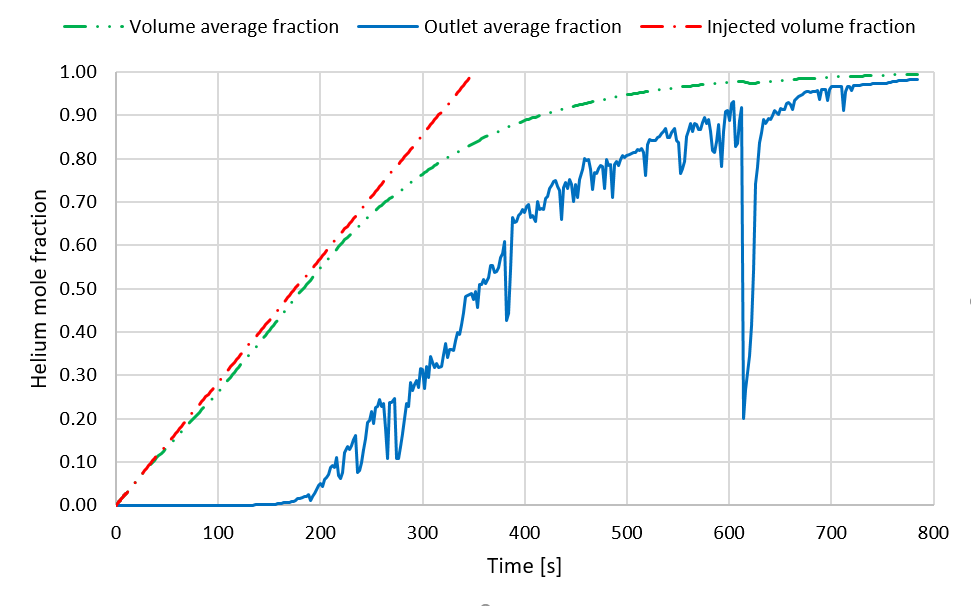}}
\caption{\label{Fig:CFD:PurProf} Helium mole fraction as function of time, during the flushing transient. The curves represent the purity averaged over the fluid volume and the outlet boundary, compared to the injection curve (normalized to the fluid volume)}
\end{figure}

As mentioned, the cryogenic purification process cannot be started until the average purity in the vessel reaches 85\%, in order to avoid condensation and freezing of air components in the cryogenic parts of the system. The purity of the gas is monitored by the passivation system at the outlet of the vessel, on the return pipeline: it is not practical to monitor directly the average purity in the vessel. Figure~\ref{Fig:CFD:PurProf} shows the helium purity evolution with time from the simulation results. The three curves represent the following quantities:
\begin{itemize}
\item Injected volume fraction: volume of pure helium injected in the vessel, normalized by the vessel free volume (74.5 m$^3$). An injected volume fraction of 1.0 indicates that an amount of pure helium equal to the vessel free volume has been pumped into the vessel. In the specific case of this simulation, the volume fraction reaches 1.0 after approximately 6 min and 2.0 after approximately 12 min; 
\item Outlet average fraction: surface average helium purity (weighted by mass) at the outlet of the helium vessel. This quantity is the purity that would be measured by an ideal purity meter positioned immediately downstream the outlet of the helium vessel. This quantity is representative of the value that would be measured by the instrumentation on the passivation system;  
\item Volume average fraction: helium purity in the helium vessel, averaged over the fluid volume. This quantity is the parameter that ideally would need to be monitored by the purification system in order to control the process.
\end{itemize}

The results show that the flushing process allows to reach a high purity without losing a large amount of helium before being able to start the purification process: the volume average fraction reaches 85\% (minimum threshold for purification process start) at slightly more than 1 volume change (Injected volume fraction equal to 1). The behavior of the volume average fraction is asymptotically increasing to 1; the average purity is 85\% at 1 volume change and higher than 98\% at 2 volume changes (\textasciitilde700 s). The outlet average fraction is delayed due to the lag generated by the helium transit time through the vessel; it consistently exceeds 90\% only towards the end of the second volume change.

\begin{figure}
\centering
\resizebox{0.8\textwidth}{!}{
\includegraphics{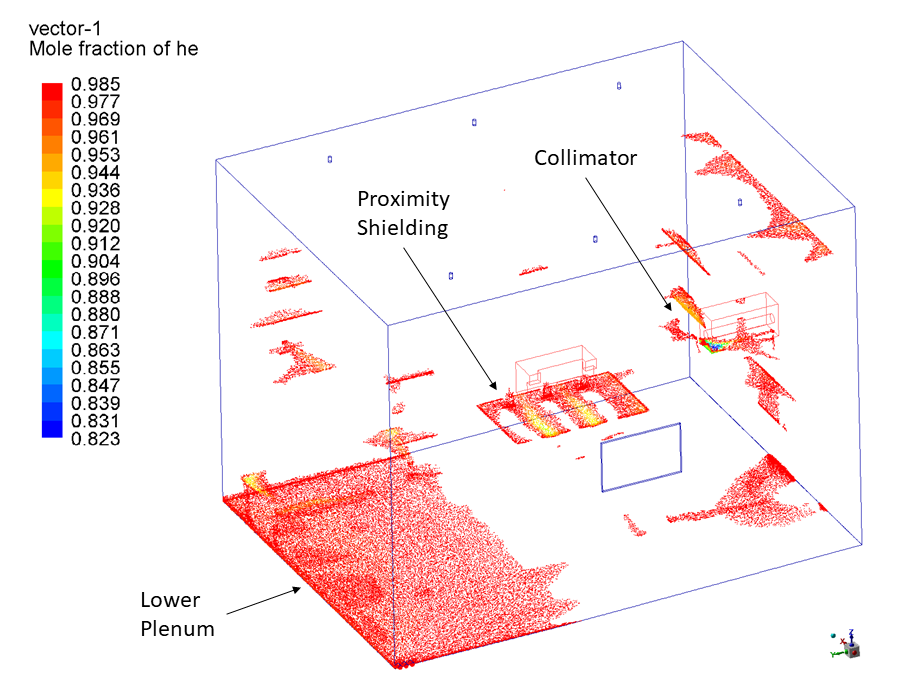}}
\caption{\label{Fig:CFD:Pockets} Vector plot of helium mole fraction at 2.23 volume changes (clipped at  98.5\%), showing areas of stagnation of air}
\end{figure}

Figure~\ref{Fig:CFD:Pockets} shows the helium purity distribution at 2.23 volume changes; the data is clipped at 98.5\% in order to highlight the areas where air concentration is higher (air pockets).
The minimum purity is about 82\% and it is likely located in the collimator area and below the target region.  
As for the steady simulation, a large pocket of air is identified at the bottom of the model, in the lower plenum; this pocket would naturally disappear for longer simulation time.

Not many low-purity areas are identified, proving that the current design and gaps layout provide good circulation patterns for helium. Before starting the helium purification process, at least 2.0 volume changes should be reached in order to avoid formation of air pockets. In terms of stagnation of impurities, horizontal gaps coupled with corners represent the most critical locations; flow-paths for helium need to be designed via precise sizing of blocks and gaps.

\section{Design of the Helium Passivation System}
\label{Sec:PSDes}

\begin{figure}
\centering
\resizebox{0.85\textwidth}{!}{
\includegraphics{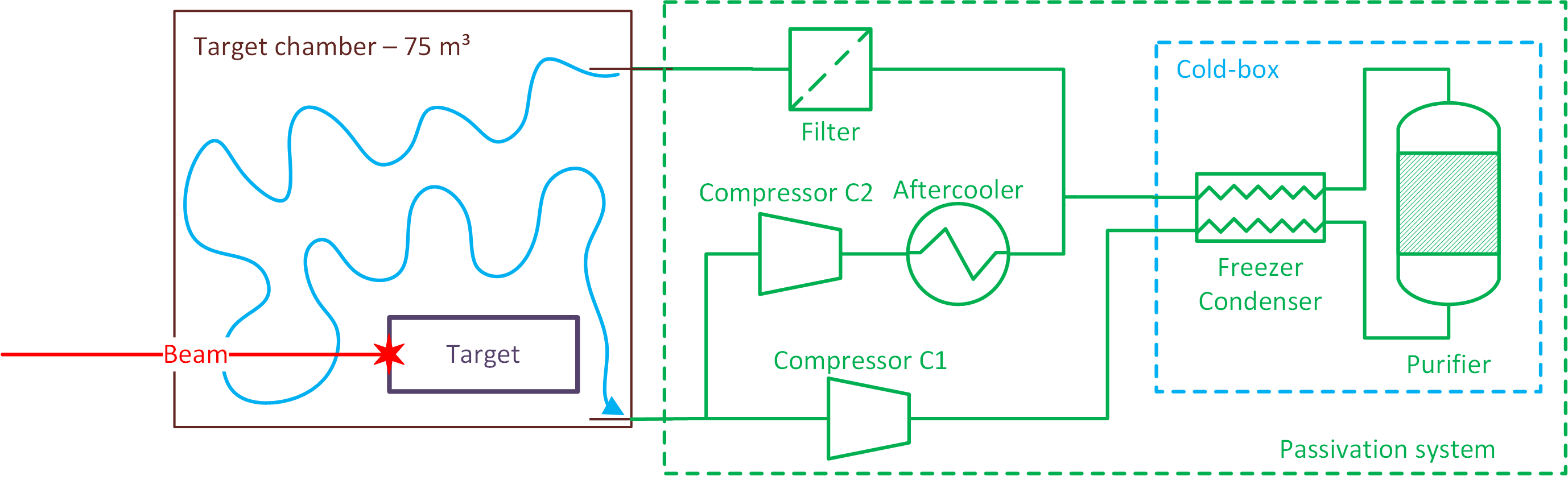}}
\caption{\label{Fig:PSDes:Intro:Sketch} Conceptual representation of the passivation system connected to the BDF Target Complex helium vessel.}
\end{figure}

CERN, in collaboration with ILK Dresden (\url{https://www.ilkdresden.de/}), has developed a detailed design of a helium cooling and passivation system for the vessel of the Beam Dump Facility. The passivation system is in charge of purifying the helium that is circulated through the vessel. Figure~\ref{Fig:PSDes:Intro:Sketch} shows a conceptual representation of the passivation system integrated with the helium vessel. The passivation system extracts the helium at the bottom of the vessel, which contains the BDF target and has a free gas volume of 75 m$^3$, purifies and cools it and pumps it back to the top of the helium vessel. The passivation system is branched in two parallel lines, responsible for the following processes:
\begin{itemize}
    \item A cooling line. This line is identified by the blower/compressor C2; it is in charge of removing 3 kW from the helium in the chamber, and keeping its temperature below 45\textdegree{}C;
    \item A purification line. This line is identified by the compressor C1, and it includes a cold-box unit. This unit is in charge of purifying helium and supplying it to the vessel at a purity level of 99.99\% (design purity);
\end{itemize}

The system is designed to supply helium at a purity of 99.99\% in order to maintain a nominal purity of 99.9\% inside the helium vessel, averaged over the volume.

\subsection{Requirements for the passivation process}
\label{Sec:PSDes:Reqs}

The system is designed to purify a volume of helium of roughly 75 m$^3$ in a worst-case scenario of 1 m$^3$/day impurity ingress into the helium vessel. This impurity ingress/production rate is cumulative for all gas components (including hydrogen and tritium from ionization and out-gassing) and it is estimated based on experience in similar facilities, the expected leak-tightness of the helium vessel and from the outgassing of the internal components, such as concrete and steel assemblies. However, it is expected that, in operational conditions, the release rate of impurities is lower than the mentioned threshold. Obtaining an accurate estimate of this parameter is relatively difficult, because it depends on the leak-tightness of the vessel and on the out-gassing of all equipment inside the vessel, particularly the shielding blocks.
Table~\ref{Tab:PSDes:Reqs} lists the main requirements for the helium passivation system.

\begin{table}[hbtp]
\centering
\caption{\label{Tab:PSDes:Reqs} Main technical requirements for the BDF target vessel passivation system.}
\begin{tabular}{|r|c|}
\hline
Vessel operating pressure {[}Pa{]} relative to external volume             & +50   \\
Vessel free volume {[}m$^3${]}                                      & 75    \\
Maximum helium return temperature {[}$^{\circ}$C{]}        & 45    \\
Cooling power {[}kW{]}        &     3    \\
Minimum vessel helium purity {[}\%{]}                               & 99.9   \\
Maximum impurity ingress {[}m$^3$/day{]}                            & 1    \\
Main impurity type {[}-{]}                                          & Air   \\ 
Number of startups per year {[}-{]}                                 & 2   \\ 
\hline
\end{tabular}
\end{table}

The system is planned to be operated based on the following three operational modes:
\begin{itemize}
    \item Preparation mode: this mode is operated during the experiment start-up phase and it lasts less than 24 h; the passivation system and its cryogenic equipment are prepared for nominal mode operation and air is replaced with helium in the vessel; 
    \item Nominal mode: this mode is operated during the main experimental phase, when the beam is on. In this mode of operation, the helium is circulated, cooled and purified by the helium passivation system;
    \item Purge mode: this mode is operated at the end of the experimental phase. Helium is purged from the vessel and replaced with air; the passivation system is deactivated.
\end{itemize}
During preparation mode helium is flushed through the vessel and purified up to the nominal purity in less than 24 hours. The process is performed in two phases:
\begin{itemize}
    \item Phase 1, flushing: pure helium is released from high-pressure cylinders into the vessel; at the same time a mixture of air and helium with increasing helium concentration is vented from the top of the vessel. When purity at the outlet of the target chamber reaches 85\%,  the mixture is ready for purification and the process switches to phase 2;
    \item Phase 2, purification: injection from cylinders is stopped and the gas mixture is circulated through the purification unit until reaching the nominal purity.
\end{itemize}
The average volume fraction curve in Figure~\ref{Fig:CFD:PurProf} illustrates the evolution of the helium purity in the chamber during the preparation process.
When the preparation of the system is completed, the nominal mode is started. In this mode, the system maintains the purity above the nominal value and provides cooling. The cooling line operates in parallel to the purification line, so the two functions can be operated simultaneously.

\subsection{Process description and system performance}
\label{Sec:PSDes:Proc}

% Proposed solution 

The passivation system design is based on the low-temperature adsorption (LTA) process, a cryogenic process performed at 77 K that allows the removal of impurities from a gas via condensation/freezing and adsorption. The basic setup for the process is shown in Figure~\ref{Fig:PSDes:Intro:Sketch} (Cold-box): the condensation/freezing is performed via the condenser-freezer and it allows the removal of high boiling point components such as moisture, CO$_2$ and oil vapors, whereas the adsorption process is performed in the purifier and it allows the removal of oxygen, nitrogen and other minor gas components. The purifier is a vessel containing adsorbing material that requires to be regenerated when it is saturated of impurities.

% System performance

The basic performance parameters of the system are listed in Table~\ref{Tab:PSDes:Proc:Perf}.
The purification process is performed at 15 bar(a), in order to maximize the adsorption capacity, and roughly 75 Nm$^3$/h, equivalent to one volume change per hour.
The protective action parameter is the operational time at constant inlet impurity level after which the adsorber material requires regeneration; at high inlet purity, regeneration is required less frequently. 

During startup phase, the purity level is not constant, so in principle the protective action time at constant purity cannot be quantified for this situation. However, since the protective action time at 15\% impurity level is 4 h, at the end of the startup purification the adsorption capacity of the adsorber is exhausted and needs to be regenerated. 

Instead, for the nominal operation scenario, a stationary condition is reached between the impurity ingress and the purification rate, resulting in an equilibrium purity of 99.95\%.
In this condition, the protective action time is more than 30 days; this is equivalent to a maximum regeneration interval of more than a month, which is compatible with the operation of the experimental area.

\begin{table}[hbtp]
\centering
\caption{\label{Tab:PSDes:Proc:Perf} Main performance parameters for the purification process of the passivation system.}
\begin{tabular}{|l|c|}
\hline
Protective action at constant 85\% purity {[}h{]}                    & 4   \\
Protective action at constant 98\% purity {[}h{]}                    & 24   \\
Protective action at constant 99.95\% purity {[}d{]}                 & 30   \\
Pressure of the adsorber {[}bar(a){]}                               & 15   \\
Purification flow rate {[}Nm$^3$/h{]}                                & 75    \\
Helium supply temperature {[}$^{\circ}\mathrm{C}${]}                & 20    \\
System leak-tightness {[}mbar*l/s{]}                                & 10$^{-6}$   \\
Liquid nitrogen consumption {[}kg/h{]}                              & 12   \\
Regeneration time {[}h{]}                                           & 6   \\
Adsorber regeneration temperature  {[}K{]}                          & 385    \\
\hline
\end{tabular}
\end{table}

% Process description

\subsubsection{Process description}

The nominal operation of the passivation system is based on two main processes running in parallel, a purification process and a cooling process. Figure~\ref{Fig:PSDes:Des:PID} shows the P\&ID of the system and provides a basic illustration of the process.
The purification process is structured according to the following steps:
\begin{enumerate}[label=\arabic*.]
    \item Impure helium at pressure of roughly 1.0 bar(a) and temperature below 45\textdegree{}C is extracted from the helium vessel;
    \item Impure helium is supplied to the compressor suction and compressed to a pressure of about 15 bar(a);
    \item Impure helium is cooled by an aftercooler;
    \item Impure helium flows through the shell space of the condenser-freezer:
        \begin{itemize}[label={--}]
            \item Helium is cooled by the backflow of purified helium and nitrogen vapor to a temperature of 88K;
            \item Impurities such as moisture, CO$_2$ and oil vapors freeze;
        \end{itemize}
    \item Impure helium is further cooled down while entering the adsorber vessels, that are immersed in a liquid nitrogen bath, to a temperature of 80K;
    \item Helium flows through the adsorber charcoal, activated carbons and molecular sieve, which adsorb the remaining gas impurities;
    \item Pure helium is directed back to the condenser-freezer for precooling the impure helium flow;
    \item Pure helium is filtered via particle filters to remove residuals from the adsorbent material;
    \item Helium at a purity level of 99.99\% is delivered to the supply line of the helium passivation system.
\end{enumerate}

The cooling process running in parallel to the purification process is designed for removing a maximum of 3 kW from the helium in the vessel, mainly coming from the shielding blocks that are heated by radiation. This process extracts helium at the bottom of the vessel, so that impurities, that are heavier than helium, are extracted from the vessel. The flow rate generated by a blower is pumped through an aftercooler and returns to the top of the helium vessel. The flow direction can be reversed via a valve manifold, in case circulation from bottom to top of the helium vessel is required (particularly for high heat load conditions, in which natural circulation effects could be relevant).

%% Purity evolution
\subsubsection{Purity evolution during start-up}

\begin{figure}
\centering
\resizebox{0.7\textwidth}{!}{
\includegraphics{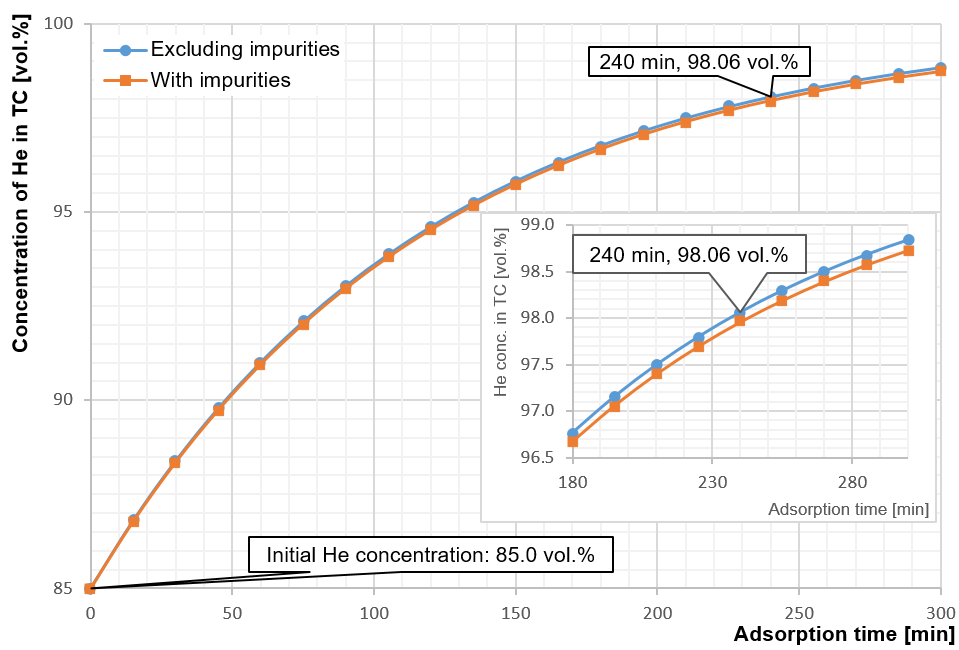}}
\caption{\label{Fig:PSDes:Proc:Pur6h} Evolution of helium purity in the helium vessel from 85 vol.\% to 98 vol.\% during initial purification. The small difference between the curves considering and not considering for the impurities implies that the effects of outgassing/leakage are negligible during the startup process.}
\end{figure}

\begin{figure}
\centering
\resizebox{0.7\textwidth}{!}{
\includegraphics{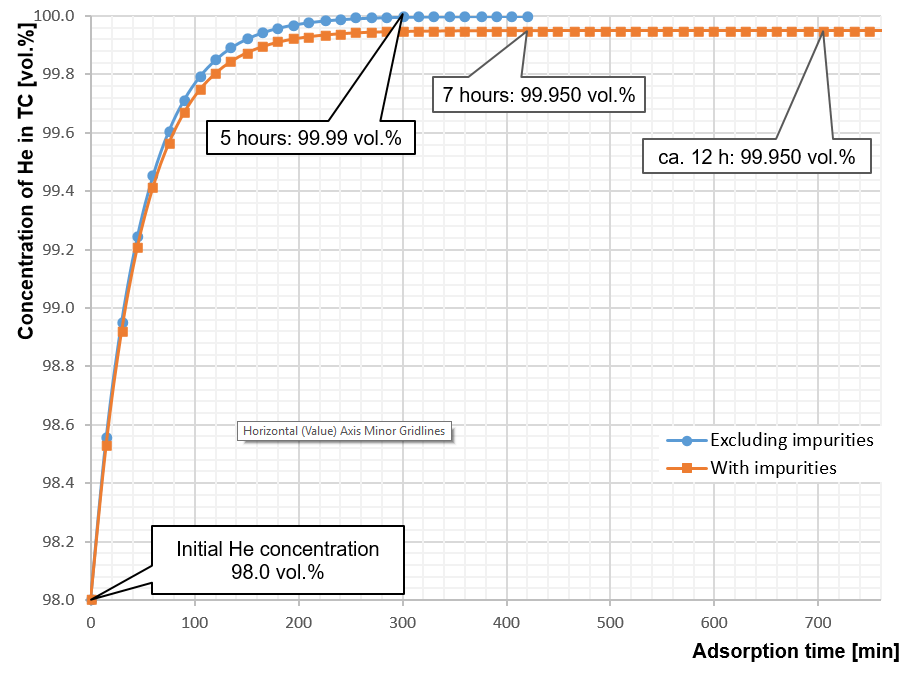}}
\caption{\label{Fig:PSDes:Proc:Pur99} Evolution of helium purity from 98 vol.\% to 99.99 vol.\% in the helium vessel, showing that the impurity ingress rate has a direct effect on the maximum achievable average purity.}
\end{figure}

After finishing the flushing operation, the helium purity inside the helium vessel is 85\% and the first purification can be performed. During the purification process, the average purity inside the helium vessel changes with time; Figure~\ref{Fig:PSDes:Proc:Pur6h} shows the purity evolution from 85\% to 98\% and Figure~\ref{Fig:PSDes:Proc:Pur99} shows the purity evolution above 98\%. The figures show the purity evolution with and without the contribution of the impurity source of 1 m$^3$/day due to outgassing and leakage.

Figure~\ref{Fig:PSDes:Proc:Pur6h} shows that the influence of the leakage on the purity evolution in the range 85-98\% is minimal: the difference in time needed for reaching 98\% between the two cases is less than 10 min for a total duration of 4 hours. Instead, in the purity region above 98\% (Figure~\ref{Fig:PSDes:Proc:Pur99}), the leakage limits the maximum achievable purity:
\begin{itemize}
\item If the leakage is not taken into account, the purification takes 5 hours for reaching an average purity of 99.99\% from 98\%;
\item If the leakage is taken into account, the process is not capable of reaching 99.99\% purity inside the helium vessel. Instead an asymptotic value is achieved, equal to 99.95\%.
\end{itemize}
The described condition is acceptable for the specific design of the helium vessel, which requires a minimum average purity of 99.9\% inside the helium vessel. If the design conditions are not achievable because of higher leak rate, and the leak tightness of the helium vessel cannot be improved, an upgrade of the compressor is necessary.

At the saturation point, the adsorber cannot capture particles and the impurities are pumped back to the helium vessel. After saturation of the adsorber, the average purity inside the helium vessel decreases at a rate of 0.056~\%/h, due to outgassing and leakage; a purity of 99.5~\% is achieved in approximately 10 hours. A 10 hours margin allows for performing a full regeneration of the adsorber.
If leakage rate is higher than expected, a redundant cold-box unit is needed in order to prevent the purity from being too low. In this condition the adsorber breakthrough is prevented via predictive analysis: an additional small capacity adsorber is added downstream the main adsorber and gas is sampled and analyzed from a connection preceding the small adsorber; whenever the main adsorber is saturated, the event is detected via sample analysis and the small adsorber provides the margin for switching to a redundant unit or taking the necessary actions.

%% Controls
\subsubsection{Process control}

The system is fully automated; control and monitoring is done via PLCs, that can be installed in a remote surface location, so that it is not affected by prompt radiation from the target complex and can be accessed independently of the beam operation status.

The helium temperature is monitored by temperature sensors on the supply and return piping and it is controlled by regulating the flow rate on the chilled water side of the aftercooler.

The pressure inside the helium vessel is monitored by a pressure differential transmitter providing the pressure difference between the helium vessel and its surrounding environment. The pressure is regulated to maintain a +50 Pa overpressure inside the vessel. The control is achieved by releasing pure helium from helium storage cylinders whenever the pressure is below +40 Pa (vessel is pressurized) and by venting helium downstream the blower when pressure is higher than +60 Pa.

The helium purity is monitored at four points along the circuit, two at the inlet and outlet of the passivation system and two at the inlet and outlet of the cold-box. The helium passivation system design foresees to use electro-chemical sensor technology for the purity meters, which allow the measurement to be performed continuously in time. A capillary pipe extracts the helium to be analyzed and delivers it to the instrument, causing a delay in measurement of a few minutes; this delay is determined by the flow rate of the gas through the capillary, the pressure of the system at the sampling location and the type of detector. 
Since the pressure level in the helium vessel does not provide a sufficient pressure head for pumping helium through the helium analyzer, it is necessary to install an oil-free flow booster to increase the pressure to the level required by the analyzer.

\subsection{Design of the helium passivation system}
\label{Sec:PSDes:Des}

% Intro and PID

\begin{sidewaysfigure}
\resizebox{1.0\textwidth}{!}{
\includegraphics{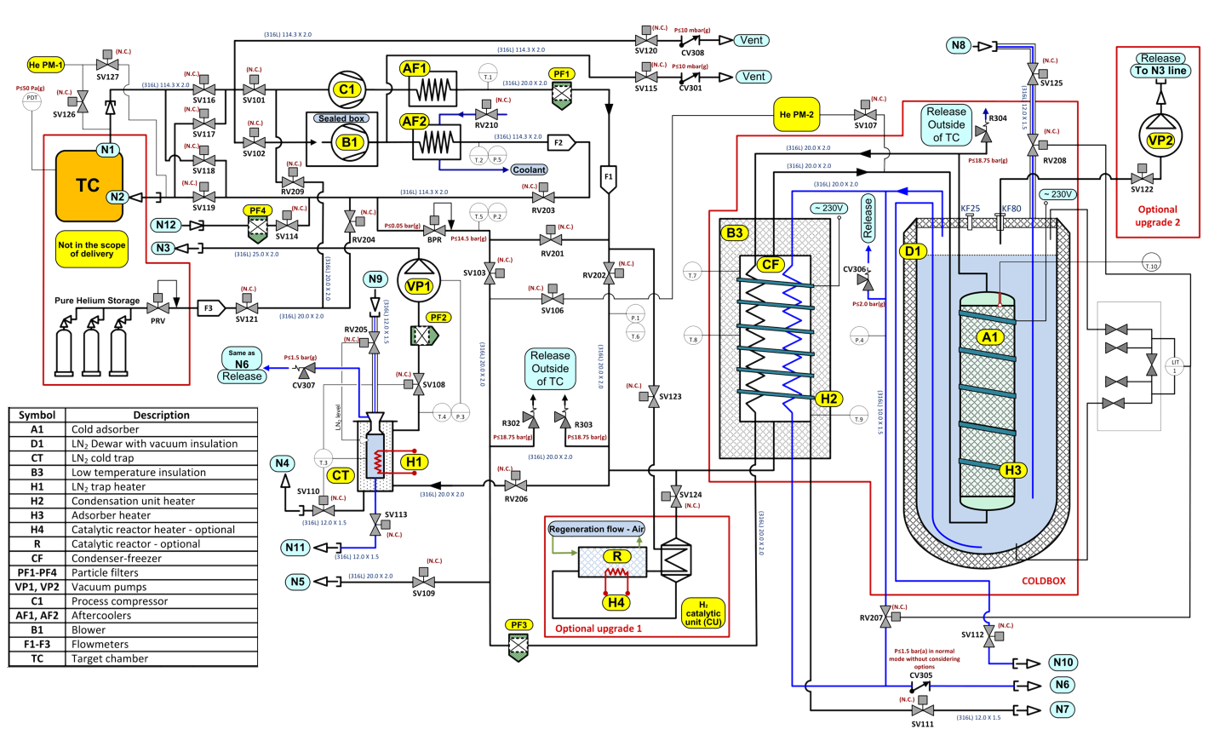}}
\caption{\label{Fig:PSDes:Des:PID} Piping and instrumentation diagram for the helium passivation system foreseen for the CERN's Beam Dump Facility (BDF) target complex.}
\end{sidewaysfigure}

The passivation system design has been developed at a detailed level for prototyping; Figure~\ref{Fig:PSDes:Des:PID} shows the preliminary piping and instrumentation diagram (P\&ID) for the passivation system.
Besides the cooling and purification functions, which have been described in section \ref{Sec:PSDes:Proc}, the system is designed so that a series of secondary processes can be performed in order to allow the correct operation of the system. The main secondary processes that need to be performed are the following:

\begin{enumerate}
    \item Start-up flushing. This process allows to replace air inside the helium vessel with helium up to a purity of 85\%. During this process, the mixture is released to the atmosphere, so part of the helium is lost. Pure helium is supplied by pressurized cylinders on the supply line, whereas the gas mixture is vented via a vent valve on the outlet line, or downstream the blower, in case some boost is required to sustain circulation. 
    \item Cold-box cooling down. This process allows to prepare the cold-box before the purification is started. The cooling-down is achieved by pumping liquid nitrogen to the cryogenic dewar containing the adsorber vessels. The condenser-freezer is also cooled down by flowing LN vapor from the dewar, before being released to the atmosphere. 
    \item Regeneration. This process is performed every time the adsorber is saturated; it lasts 6 hours and it is performed at 385 K (Table~\ref{Tab:PSDes:Proc:Perf}). During this process, liquid nitrogen is drained from the dewar and heaters are turned on to release the impurities from the adsorber. A vacuum pump extracts the gases from the adsorber and condenser freezer; a cold trap protects the vacuum pump from high condensation point components. After vacuuming, the cold-box is filled with helium, ready for the next purification phase.
    \item Purging. During this phase, helium from the top of the helium vessel is vented to the atmosphere, and it is replaced by filtered air entering the bottom of the helium vessel via a dedicated inlet pipe.
\end{enumerate}

%\begin{figure*}
%\end{figure*} 

% 3D model

\begin{figure}
\centering
\resizebox{0.98\textwidth}{!}{
\includegraphics{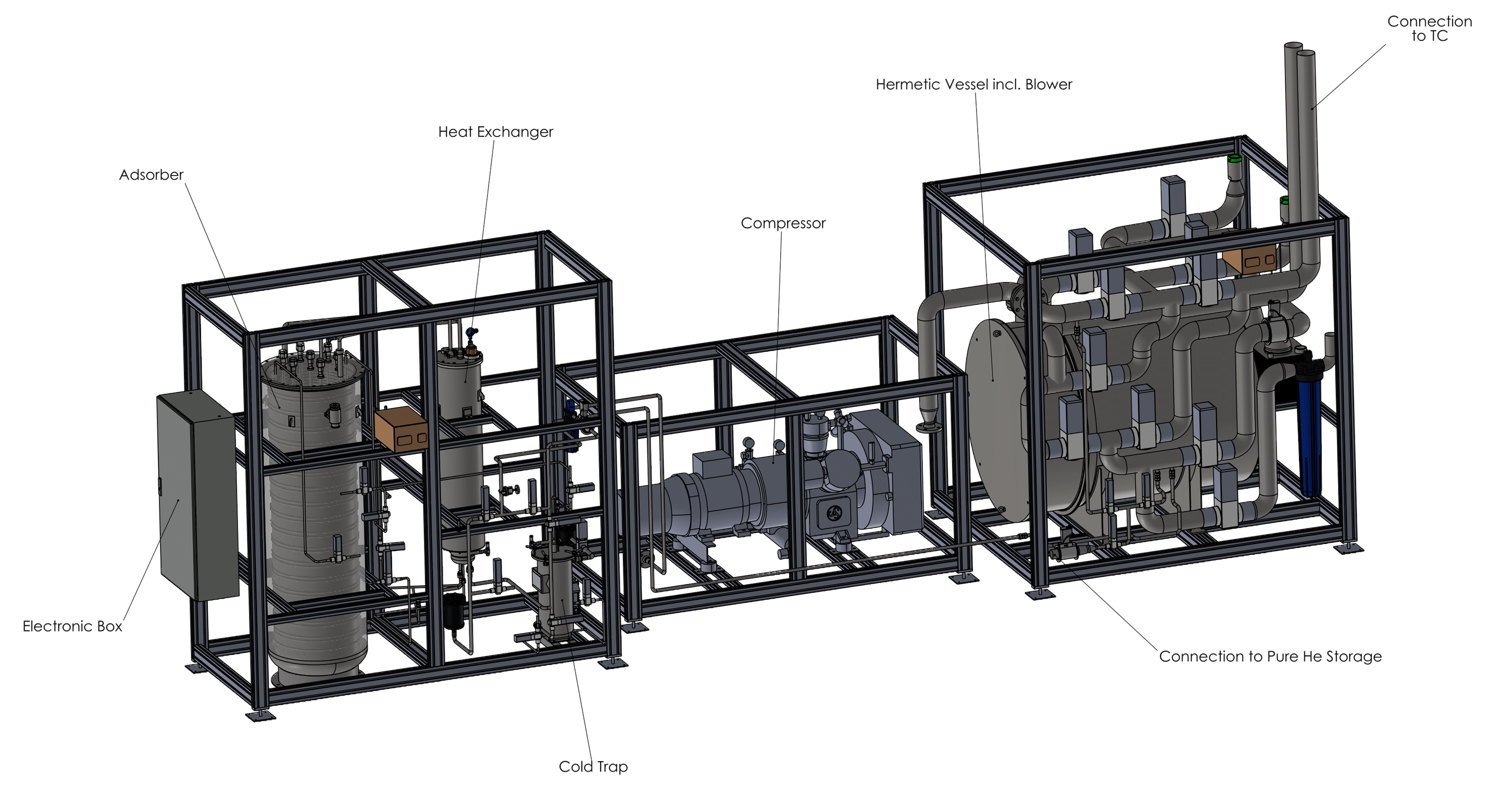}}
\caption{\label{Fig:PSDes:Des:3D} 3D model of the helium passivation system, showing the different components of the assembly.}
\end{figure}

Figure~\ref{Fig:PSDes:Des:3D} shows the preliminary integration of the system and its components. The system is composed of three main blocks: cold-box unit, purification compressor unit and blower unit (from left to right, respectively); in terms of size of equipment, considering the cold-box unit as a reference, the indicative height of the skid is 2.5 m. 

% Explain components

The core of the system is the cold-box, of which the main components are the condenser-freezer and the adsorber vessel.

The condenser-freezer (CF in Figure~\ref{Fig:PSDes:Des:PID}) is designed for condensing and freezing moisture vapor, carbon dioxide (CO$_2$) and traces of hydrocarbons. The CF significantly reduces the purification load on the adsorber and increases its lifetime. The main design parameters of CF are shown in Table~\ref{Tab:PSDes:Des:CF}.
The CF is a three-stream heat recuperator with impure helium on the shell side and purified helium and nitrogen vapors on the two tube sides, made of twisted copper ribbed pipes supported around a stainless steel rod. The CF is provided with NiCr electric heating tape for the regeneration procedure, and the cryogenic part of the heat exchanger is provided with perlite insulation.

\begin{table}[hbtp]
\centering
\caption{\label{Tab:PSDes:Des:CF} Main design parameters of the condenser-freezer heat exchanger.}
\begin{tabular}{|l|c|}
\hline
Unit heat load {[}kW{]}                         & 3.6   \\
LMTD {[}K{]}                                    & 6.2   \\
Effective surface area {[}m\textsuperscript{2}{]}                 & 11   \\
Surface area margin {[}\%{]}                    & 50   \\
\hline
\end{tabular}
\end{table}

The adsorber vessel is designed as a bundle of 7 cylindrical stainless-steel vessels connected in series and filled with a mixture of adsorbent material. Each vessel has an outer diameter of 180 mm and a height of 1400 mm for a total volume of approximately 250 liters. All components of the vessels are welded. NiCr heater tape is provided for each vessel for the regeneration of the adsorber. The bundle of vessels is designed to operate at 15 bar and 75 m$^3$/h nominal flow.
The adsorber is connected in series with the CF and they work simultaneously. 
The impure helium flowing through the adsorber reaches the design purity in a single pass. The composition of the mixed adsorbent material is determined by considering the adsorption capacity and selectivity of the adsorbents for each component of the gas mixture; since the expected main impurity is expected to be air, the mixed adsorbent requires consistent depletion capacity for two major air components, nitrogen and oxygen. Based on these considerations, the design foresees 5 vessels filled with an adsorbent equivalent to 207C (90 kg), a coconut based steam activated carbon particularly selective for nitrogen, and 2 vessels filled with an adsorber equivalent to CMS H2 55/2 (35 kg), a carbon molecular sieve particularly selective for oxygen.

For increasing the sorption capacity, the adsorber vessels are submerged into a liquid nitrogen bath, contained in the liquid nitrogen dewar (D1 in Figure \ref{Fig:PSDes:Des:PID}). The dewar is a cylindrical vacuum insulated vessel at a nominal vacuum level of 10$^{-3}$ to 10$^{-4}$ mbar(a). The insulation system is constituted by a vertical double-walled multi-layer insulation. The level monitoring inside the liquid nitrogen container is performed via differential pressure sensor.  

The purification process is based on a 15 bar hermetic compressor with an adjustable flow rate in the range 54-83 Nm$^3$/h. The compressor is hermetic and gas-tight (magnetic coupling), so that no oil traces are released in the purification cycle. 
The compressor is operated by regulation of the suction pressure via variable-frequency drive.
Due to the heating of the helium during the compression phase, the compressor is provided of four cylinders and four stages; at each stage the gas is cooled via air cooling.

A blower provides the continuous circulation required for cooling at a flow rate of approximately 770 m$^3$/h. The blower needs to be gas-tight; due to the cost of a high flow-rate hermetic compressor, an alternative design has been selected, based on encapsulating a conventional blower inside a hermetic vessel. The vessel will serve as suction chamber for the intake of the blower.
The selected blower is a standard side-channel single-stage dynamic blower with contact-less rotating impeller, provided with silencers at the inlet and outlet. A brazed-plate heat exchanger is installed downstream the blower to remove the heat generated in the helium vessel and in the blower.

With respect to tritium production in the helium vessel, the helium passivation system has been designed with an option for a catalytic converter unit that is capable of converting tritium from the gas mixture (in the form of HT) to tritiated water (namely as $^{3}$HOH, also HTO). Tritiated water is then frozen in the condenser-freezer component, before entering the adsorber unit. Dedicated discharge points are provided as part of the system in order to allow appropriate disposal of waste, including tritiated water, during the regeneration procedure.

\section{Conclusions and future work}
\label{Sec:Concl}

In the framework of the Beam Dump Facility (CERN) project a detailed design of the helium passivation methodology for the target chamber has been developed. 

A series of CFD simulations has been performed to demonstrate that the current design of the target vessel allows for proper circulation of helium and prevents the formation of air pockets and impurities that could potentially compromise the operability of the facility. The analysis shows that the shielding blocks inside the helium vessel need to be designed in such a way that helium is allowed to flow through gaps to remove activated gas impurities. Further analyses will be performed, with particular focus on the effects of the heat loads during operation; the possibility of using the helium passivation system for actively cooling some of the components contained in the helium vessel will also be investigated. 

A design for the system that circulates and purifies the helium has been completed, showing the viability of the passivation of the BDF target chamber. The design has been developed at a detailed level, laying the foundations for full-scale prototyping and testing, currently foreseen in the Technical Design Phase of the installation.

\acknowledgments

The authors express their thanks to the Physics Beyond Colliders Study Group (specifically the Beam Dump Facility Project group) for the support and resources provided in the execution of the activity, as well as the EN-CV and EN-STI Groups.

% \paragraph{Note added.} This is also a good position for notes added after the paper has been written.

\newpage
\bibliographystyle{JHEP}
\bibliography{references}

% % We suggest to always provide author, title and journal data:
% % in short all the informations that clearly identify a document.
 
% \begin{thebibliography}{99}

% \bibitem{a}
% Author, \emph{Title}, \emph{J. Abbrev.} {\bf vol} (year) pg.

% \bibitem{b}
% Author, \emph{Title},
% arxiv:1234.5678.

% \bibitem{c}
% Author, \emph{Title},
% Publisher (year).

% % Please avoid comments such as "For a review'', "For some examples",
% % "and references therein" or move them in the text. In general,
% % please leave only references in the bibliography and move all
% % accessory text in footnotes.

% % Also, please have only one work for each \bibitem.

% \end{thebibliography}

\end{document}